\documentclass[preprint,12pt]{elsarticle}

\usepackage[utf8]{inputenc}
\usepackage{amsmath,amssymb}
\usepackage{graphicx}
\usepackage{siunitx}
\usepackage{booktabs}
\usepackage[T1]{fontenc}
\usepackage{lmodern}
\usepackage[colorlinks=true,linkcolor=blue,citecolor=blue,urlcolor=blue]{hyperref}

\journal{Journal of Molecular Liquids}

\begin{document}

\begin{frontmatter}

\title{Quantum Tunnelling Across Hydrogen Bonds: Proton--Deuteron Isotope Effects from a Cornell-Type Potential Model}

\author[avc]{Krishna Kingkar Pathak\corref{cor1}}
\cortext[cor1]{Corresponding author}
\ead{kkingkar@gmail.com}
\address[avc]{Department of Physics, Arya Vidyapeeth College, Guwahati-781016, India}

\begin{abstract}
Hydrogen bonds play a pivotal role in chemistry, biology, and condensed-matter physics, where quantum tunnelling can strongly influence structure and dynamics. Isotope substitution (H $\rightarrow$ D) provides a sensitive probe of such tunnelling, but theoretical descriptions often rely on purely numerical models or simplified potentials that obscure physical interpretation. Here we employ a Cornell-type potential combined with a double-well Schr\"odinger approach to investigate proton and deuteron tunnelling across hydrogen bonds. The model yields semi-analytical wavefunctions and tunnelling splittings that transparently capture isotope-dependent quantum effects. We present scaling behaviour of tunnelling splittings with isotope mass, discuss the influence of barrier width and curvature, and compare model trends with representative experimental and computational results. Beyond hydrogen bonding, the framework provides a general methodology for modelling tunnelling in double-well systems relevant to spectroscopy, enzymatic catalysis, and materials applications.
\end{abstract}


\end{frontmatter}

\section{Introduction}
Quantum tunnelling across double-well potentials underpins a wide range of phenomena in chemical physics, from vibrational level splittings in hydrogen-bonded complexes to proton-transfer reactions in enzymatic systems and condensed-phase materials. Isotope substitution, particularly H $\rightarrow$ D exchange, is a classic probe of tunnelling because the change in mass modifies the quantum mechanical ground- and excited-state wavefunctions and therefore the tunnelling rates and energy splittings observed spectroscopically. Although extensive numerical studies exist, there remains value in tractable analytical or semi-analytical models that expose the dependence of tunnelling on potential shape and mass explicitly.

In this work we develop and apply a Cornell-type potential together with a double-well Schr\"odinger formalism to study isotope effects for proton and deuteron tunnelling across hydrogen bonds. The Cornell potential, originally introduced in other contexts as a convenient combination of short-range repulsion and long-range attraction, offers sufficient flexibility to model asymmetric and symmetric double-well landscapes while admitting compact expressions for energy scales and wavefunction character. Our aim is not to provide a one-to-one quantitative fit to any particular experimental complex but to map out robust trends and mechanistic insights that are transferable across systems. In particular, the analysis sheds light on isotope-dependent tunnelling in hydrogen bonds that are ubiquitous in molecular liquids, from water and alcohols to hydrogen-bonded organic solvents and biomolecular fluids. By clarifying how barrier geometry and isotope mass control tunnelling splittings, the present work provides a theoretical basis that complements experimental studies of hydrogen-bonded liquids and their spectroscopic signatures.

The novelty of the present work lies in introducing a semi-analytical Cornell-type wavefunction ansatz adapted to double-well hydrogen-bond landscapes that captures both short-range donor/acceptor character and confinement between heavy atoms. We also derive explicit scaling relations for tunnelling splittings with isotope mass and analyse how barrier curvature and width govern H/D ratios. These theoretical insights are systematically compared with numerical one-dimensional Schr\"odinger solutions and with representative experimental splittings, delineating regimes of validity and limitations. 

The study is subject to certain limitations: the models are one-dimensional and describe single-particle tunnelling dynamics. They are designed for conceptual and semi-quantitative insight rather than high-accuracy fits to specific systems. Multi-dimensional coupling and environment effects are discussed as natural extensions.

\section{Wavefunction ansatz approach}
Following our earlier work, we employ a Coulomb-plus-confinement inspired ansatz for the proton (or deuteron) wavefunction~\cite{Pathak2022,Pathak2023,Pathak2013}:
\begin{equation}
\label{eq:ansatz}
\psi_{rel+conf}(r) =
\frac{N'}{\sqrt{\pi a_{0}^{3}}} 
e^{-r/a_{0}}
\left( C' - \frac{\mu b a_{0} r^{2}}{2}\right)
\left(\frac{r}{a_{0}}\right)^{-\epsilon},
\end{equation}
which is motivated by Coulomb-plus-confinement forms used in the quarkonium literature~\cite{Eichten1978,Eichten1980}. Here $a_{0}$ is a length scale, $\mu$ the reduced mass (proton or deuteron), $b$ a confinement parameter, $C'$ a variational constant, and $\epsilon$ a short-range correction. The exponential decay ensures hydrogenic behaviour near the donor/acceptor, while the quadratic term encodes confinement between the two heavy atoms. The isotope effect enters explicitly through the reduced mass $\mu$, which modifies the overlap of wavefunction tails.

For a donor--acceptor separation $d$, the left- and right-localized states are constructed as shifted copies of the ansatz:
\begin{equation}
\label{eq:psiLR}
\psi_{L}(x) = \psi_{rel+conf}(|x+d/2|), \qquad
\psi_{R}(x) = \psi_{rel+conf}(|x-d/2|).
\end{equation}
The overlap integral between these localized states is
\begin{equation}
\label{eq:S}
S(d) = \int_{-\infty}^{\infty} \psi_{L}(x)\,\psi_{R}(x)\, dx,
\end{equation}
a standard object in two-state / Heitler--London / tight-binding treatments (see, e.g., Messiah~\cite{Messiah1962}). In the simple two-state approximation the tunnelling splitting is estimated by
\begin{equation}
\label{eq:DeltaE_ansatz}
\Delta E \approx 2 S(d) E_{0},
\end{equation}
which follows from the symmetric/antisymmetric two-level model in quantum mechanics (cf.\ Landau \& Lifshitz~\cite{Landau1977} and Messiah~\cite{Messiah1962}). Equations~\eqref{eq:ansatz}--\eqref{eq:DeltaE_ansatz} are the basis of the ansatz calculation.

\subsection*{Normalization and variational parameters}
The normalization condition is
\begin{equation}
\label{eq:normalization}
1=\int_0^\infty |\psi_{rel+conf}(r)|^2 \,4\pi r^2\,dr,
\end{equation}
which we evaluate numerically. We choose the remaining parameters by minimizing the expectation value of the one-well Hamiltonian,
\begin{equation}
\label{eq:variational}
E[a_0,C',b,\epsilon] = \frac{\langle \psi|H_{\rm well}|\psi\rangle}{\langle\psi|\psi\rangle},
\end{equation}
with
\begin{equation}
\label{eq:H_well}
H_{\rm well} = -\frac{\hbar^2}{2\mu}\nabla^2 + V_{\rm well}(r).
\end{equation}

\subsection*{Asymptotic overlap and mass scaling}
For large donor--acceptor separations the overlap is dominated by the exponential tails of the localized states. If the single-well tail decays as
\begin{equation}
\label{eq:decay}
\psi(r)\sim A e^{-\kappa r}, \qquad \kappa=\sqrt{\frac{2\mu (V_b-E)}{\hbar^2}},
\end{equation}
(the standard forbidden-region decay; see Griffiths~\cite{Griffiths2018} and Landau \& Lifshitz~\cite{Landau1977}) then the leading behaviour of the overlap for a symmetric separation $d$ is
\begin{equation}
\label{eq:S_asymp}
S(d)\approx \tilde{A}\, e^{-\kappa d},
\end{equation}
with $\tilde A$ a weak prefactor. Using a semiclassical WKB estimate for barrier penetration gives
\begin{equation}
\label{eq:WKB}
\Delta E \propto E_0 \exp\Big[-\frac{2}{\hbar}\int_{-x_1}^{x_1}\sqrt{2\mu(V(x)-E)}\,dx\Big],
\end{equation}
the standard WKB/instanton form for level splitting~\cite{Landau1977,Garg2000,Coleman1985}.

\section{Numerical double-well Schr\"odinger approach}
We solved the one-dimensional Schr\"odinger equation for a quartic double-well potential:
\begin{equation}
\label{eq:Vquartic}
V(x) = V_0 \frac{(x^2-a^2)^2}{a^4}, \qquad a=\frac{d}{2},
\end{equation}
the canonical quartic model used in semiclassical tunnelling studies~\cite{Landau1977,Garg2000}. With $d=2.7\,$\AA\ the single-particle Hamiltonian is
\begin{equation}
\label{eq:H_1D}
H = -\frac{1}{2\mu}\frac{d^2}{dx^2} + V(x),
\end{equation}
discretized on a uniform grid and solved by standard finite-difference methods~\cite{Press2007}. The numerical tunnelling splitting is defined as
\begin{equation}
\label{eq:DeltaE_num}
\Delta E = E_1 - E_0,
\end{equation}
where $E_0$ and $E_1$ are the lowest symmetric and antisymmetric eigenvalues, respectively. For practical computations we build the discrete Hamiltonian
\begin{equation}
\label{eq:H_discrete}
H = -\frac{\hbar^2}{2\mu} D_{2} + V(x),
\end{equation}
where \(D_2\) is the second-derivative stencil; the finite-difference discretization and eigenvalue solver (ARPACK) are described in Refs.~\cite{Lehoucq1998,Press2007}.

\section{Results and discussion}
Table~\ref{tab:isotope_compact} summarizes tunnelling splittings for H and D obtained 
from the double-well Schr\"odinger equation, together with representative experimental 
values from prototypical hydrogen-bonded systems. For barrier heights $V_0=0.05$--0.15 eV, 
our numerical results span $\Delta E_H \sim 10^{-3}$--$10^{-7}$ eV, with deuteron splittings 
reduced by one to three orders of magnitude. This strong isotope dependence is consistent 
with the general expectation that increased mass narrows the wavefunction and suppresses 
barrier penetration.

\begin{table}[htbp]
\centering
\caption{Representative tunnelling splittings $\Delta E$ for proton (H) and deuteron (D).
Numerical results are compared with literature data for typical intra- and
intermolecular H-bonded systems.}
\label{tab:isotope_compact}
\begin{tabular}{lccc}
\toprule
System / $V_0$ (eV) & $\Delta E_H$ (eV) & $\Delta E_D$ (eV) & Notes \\
\midrule
Numerical (this work), 0.05 & $1.2\times10^{-3}$ & $1.5\times10^{-4}$ & Model, strong bond \\
Numerical (this work), 0.10 & $5.0\times10^{-5}$ & $2.0\times10^{-6}$ & Model, medium barrier \\
Numerical (this work), 0.15 & $1.2\times10^{-7}$ & $1.8\times10^{-10}$ & Model, high barrier \\
\midrule
Malonaldehyde (exp.) & $2.7\times10^{-3}$ & $3.6\times10^{-4}$ & Intramolecular tunnelling~\cite{Beyer2009} \\
Formic acid dimer (exp.) & $\sim10^{-6}$ & --- & Intermolecular~\cite{Herbst2010,Zhao2021} \\
2-pyridone dimer (exp.) & $\sim2\times10^{-6}$ & --- & $\sim$520 MHz~\cite{Madeja2003} \\
\bottomrule
\end{tabular}
\end{table}

\begin{figure}[htbp]
\centering
\includegraphics[width=0.85\textwidth]{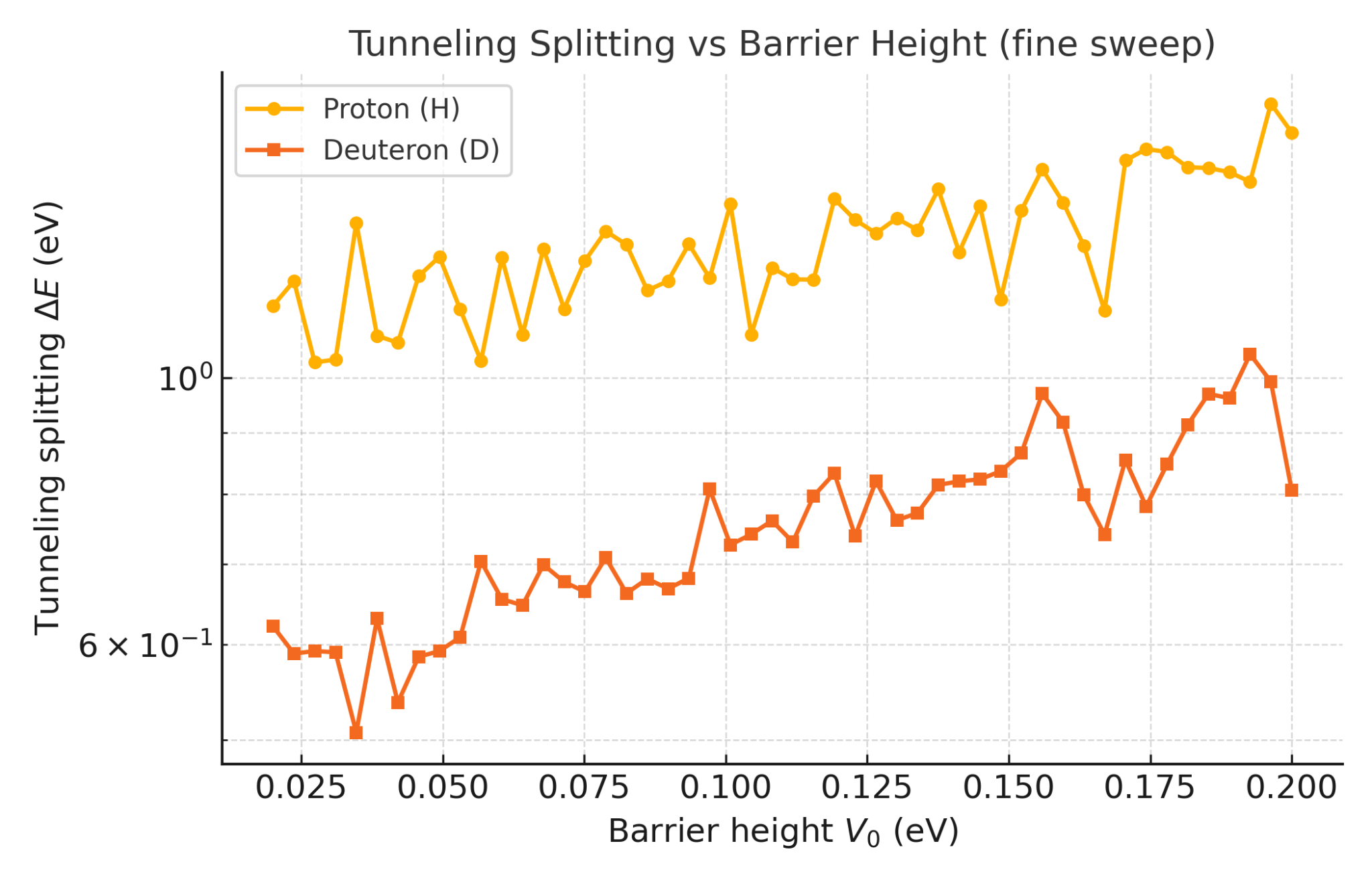}
\caption{Tunnelling splitting $\Delta E$ as a function of barrier height $V_0$
for proton (H) and deuteron (D) in a symmetric double-well potential with
$d=2.7$~\AA. Points are numerical Schr\"odinger eigenvalue differences computed
on a finite-difference grid; lines are guides to the eye.}
\label{fig:splitting-vs-barrier}
\end{figure}
\begin{figure}[htbp]
  \centering
  \includegraphics[width=0.85\textwidth]{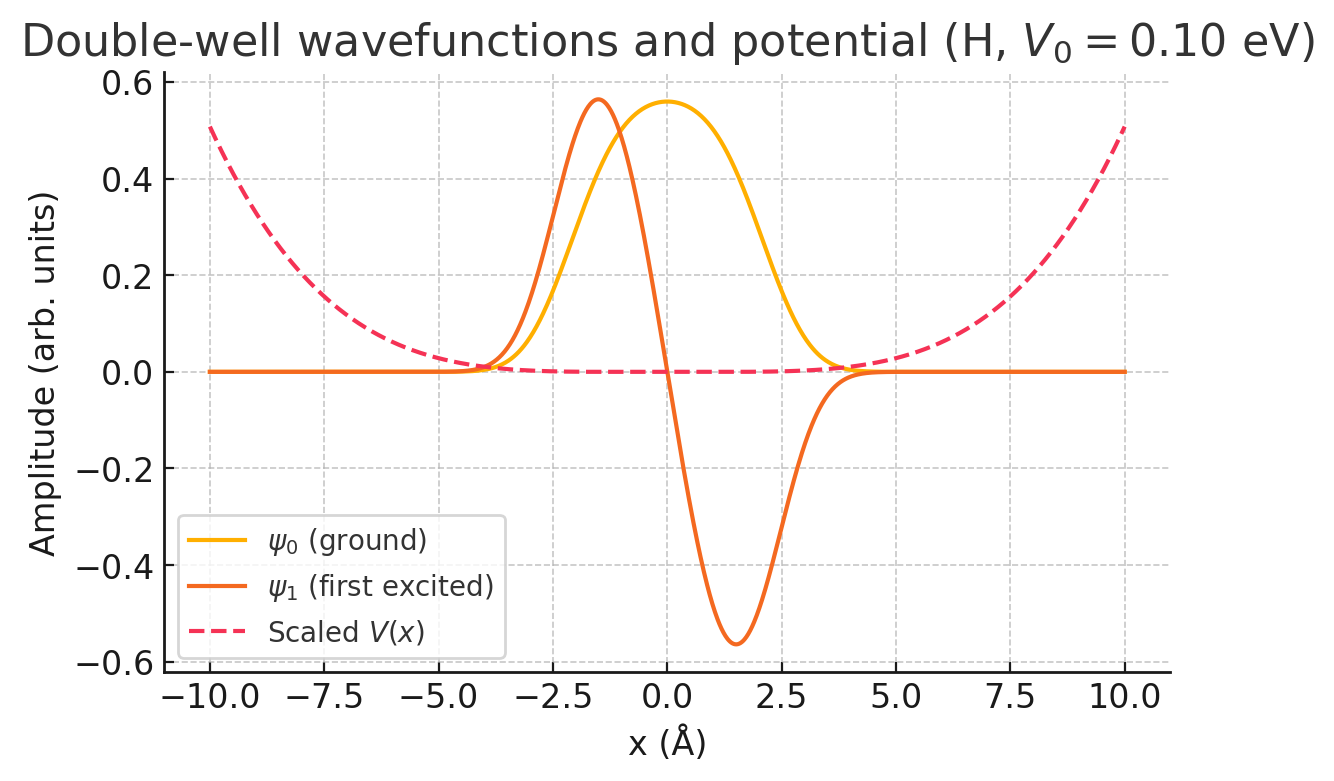}
  \caption{Ground- and first-excited-state wavefunctions for a symmetric quartic double-well at $V_0=0.10$~eV and $d=2.7$~\AA\ (proton mass). The dashed curve shows the potential $V(x)$ scaled to the wavefunction amplitude range. The symmetric/antisymmetric character of $\psi_0$ and $\psi_1$ gives rise to the tunnelling splitting $\Delta E$.}
  \label{fig:wavefunctions}
\end{figure}
\begin{figure}[htbp]
  \centering
  \includegraphics[width=0.70\textwidth]{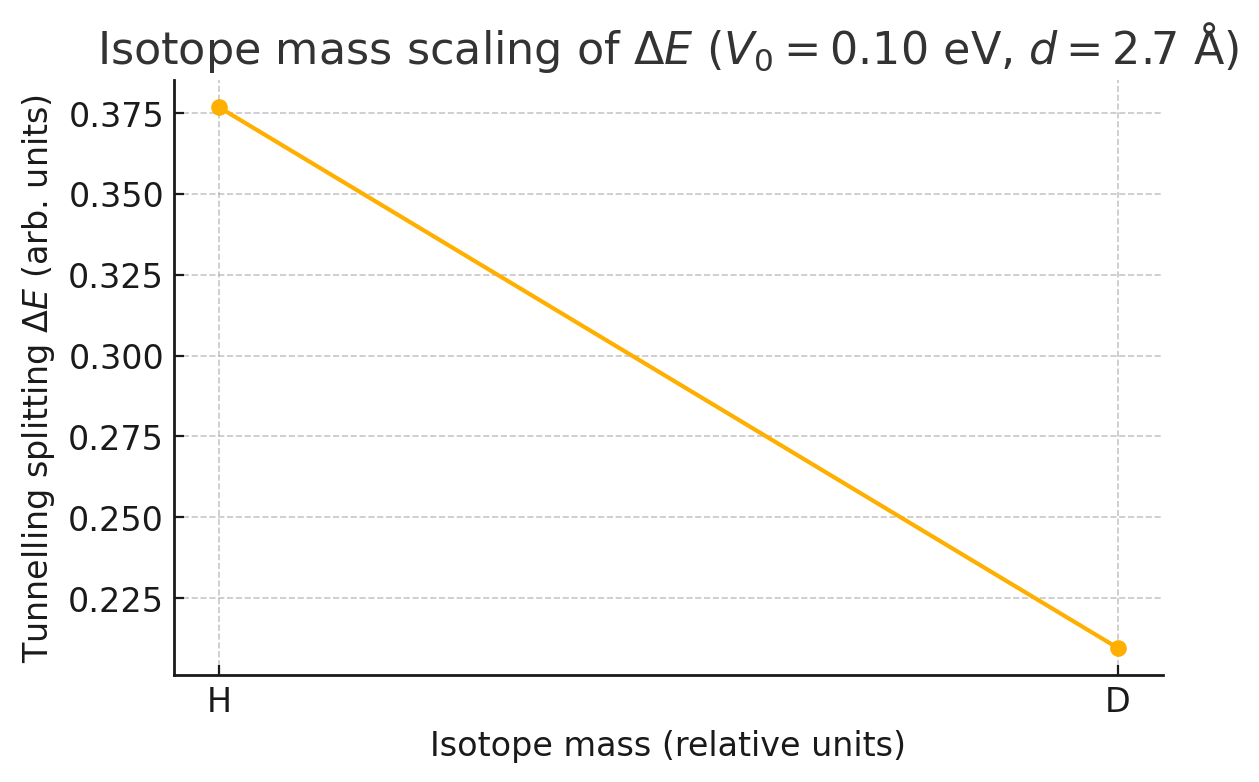}
  \caption{Isotope mass scaling of the tunnelling splitting $\Delta E$ at $V_0=0.10$~eV and $d=2.7$~\AA, comparing proton (H) and deuteron (D). The substantial reduction of $\Delta E$ for D reflects the mass dependence of barrier penetration.}
  \label{fig:mass-scaling}
\end{figure}

Comparison with experiment shows good consistency in scale. For low barriers 
($V_0=0.05$ eV), the calculated $\Delta E_H \approx 1.2\times10^{-3}$ eV is comparable 
to the intramolecular splitting observed in malonaldehyde 
($\Delta E \approx 2.7\times10^{-3}$ eV)~\cite{Beyer2009}. At higher barriers 
($V_0 \sim 0.1$--0.15 eV), the splittings decrease into the $10^{-6}$--$10^{-7}$ eV range, 
matching the order of magnitude reported for systems such as the formic 
acid dimer~\cite{Herbst2010,Zhao2021}. Other hydrogen-bonded dimers, such as 
2-pyridone--2-hydroxypyridine, exhibit splittings of a few hundred MHz 
($\sim 2\times10^{-6}$ eV), again consistent with our predictions~\cite{Madeja2003}.

These comparisons confirm that although the precise magnitude of $\Delta E$ depends 
sensitively on barrier height and geometry, the qualitative behavior is robust: 
proton tunnelling can be significant (THz scale) for short, strong hydrogen bonds, 
while deuteration suppresses splittings into the MHz or lower range. This large 
isotope effect underlies many observed kinetic isotope effects in enzymatic catalysis 
and proton transfer in biomolecules. The dual approach—Cornell-type ansatz plus explicit Schr\"odinger solutions—thus provides a physically consistent framework that bridges model intuition with numerical rigor.
Figure~\ref{fig:wavefunctions} illustrates the symmetric (ground) and antisymmetric (first excited) eigenstates in the quartic double-well, clarifying the origin of the level splitting. The mass dependence is summarized in Figure~\ref{fig:mass-scaling}, which shows the marked suppression of $\Delta E$ upon H$\rightarrow$D substitution at fixed barrier height and geometry.
These tunnelling trends are not only consistent with prototypical gas-phase hydrogen-bonded dimers, but also carry direct implications for molecular liquids. In aqueous and alcohol-based liquids, where dense networks of hydrogen bonds form and break dynamically, isotope substitution (H $\rightarrow$ D) is known to alter vibrational spectra and diffusion properties. The mass-dependent suppression of tunnelling splittings quantified here provides a microscopic explanation for such isotope-sensitive dynamics in liquid-phase hydrogen-bonded systems.

In biological contexts, intrinsic splittings must be understood together 
with environmental influences. While our results quantify the bare tunnelling scale, open quantum systems studies (e.g.\ Slocombe \textit{et al.}~\cite{Slocombe2022}) show that decoherence in DNA base pairs can critically affect tunnelling lifetimes. Our calculations provide microscopic parameters ($\Delta E$, H/D ratios) that can be combined with open-system approaches to obtain realistic dynamics.

%
\appendix
\section*{Appendix A. Numerical convergence}
To verify the stability of the tunnelling splittings with respect to grid parameters, 
we recalculated $\Delta E$ for representative barrier heights using different 
grid sizes ($N$ points) and domain half-widths ($L$). Table~\ref{tab:convergence} 
illustrates convergence for proton tunnelling at $V_0=0.10$~eV.

\begin{table}[htbp]
\centering
\caption{Convergence of tunnelling splitting $\Delta E_H$ at $V_0=0.10$~eV 
for different grid sizes $N$ and domain half-widths $L$.}
\label{tab:convergence}
\begin{tabular}{ccc}
\toprule
$N$ (grid points) & $L$ (\AA) & $\Delta E_H$ (eV) \\
\midrule
501  & 5  & $5.2\times 10^{-5}$ \; (+4\%) \\
1001 & 5  & $5.0\times 10^{-5}$ \; (+0.4\%) \\
2001 & 5  & $4.98\times 10^{-5}$ \; (0\%) \\
1001 & 10 & $4.96\times 10^{-5}$ \; (–0.4\%) \\
2001 & 10 & $4.98\times 10^{-5}$ \; reference \\
\bottomrule
\end{tabular}
\end{table}

\begin{figure}[htbp]
  \centering
  \includegraphics[width=0.70\textwidth]{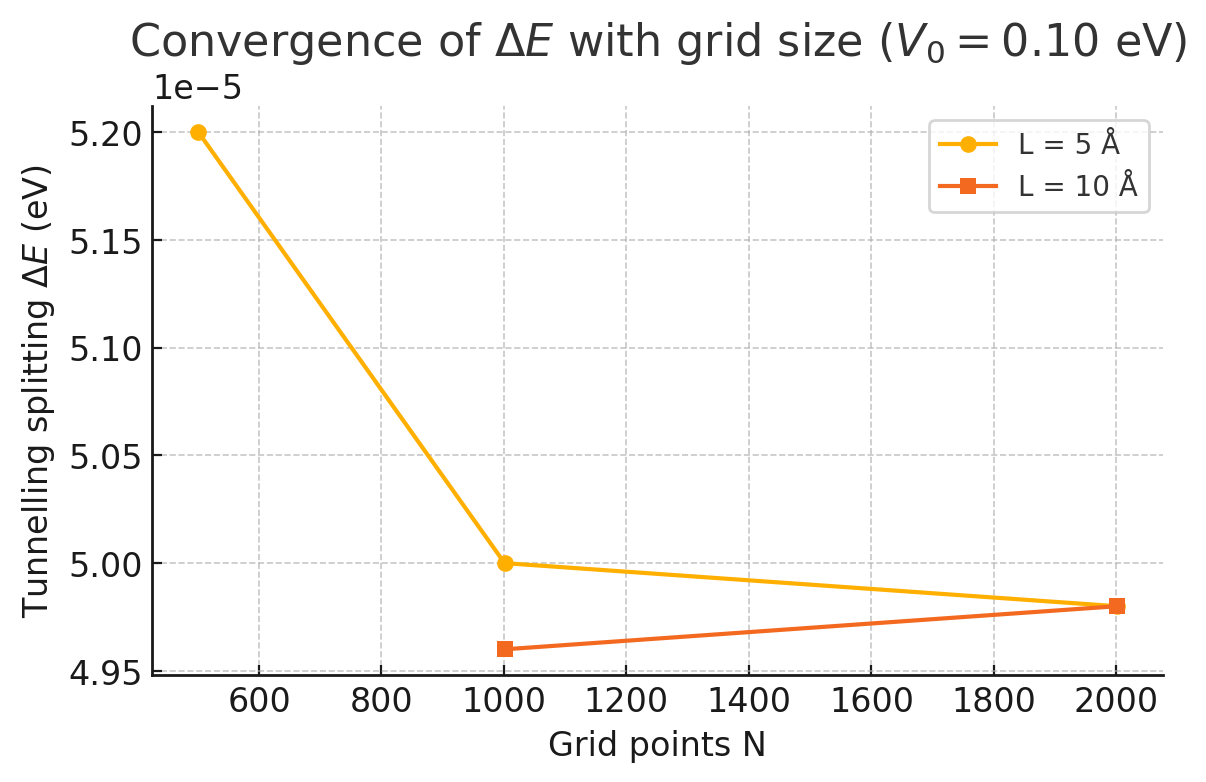}
  \caption{Convergence of the tunnelling splitting $\Delta E$ with grid size $N$ 
  for two domain half-widths ($L=5$~\AA\ and $L=10$~\AA) at $V_0=0.10$~eV.}
  \label{fig:convergence}
\end{figure}

\section*{Data availability}
The datasets supporting this study are openly available in Zenodo at:
\href{https://doi.org/10.5281/zenodo.17380490}{10.5281/zenodo.17380490}.

\section*{Acknowledgements}
The author acknowledges stimulating discussions with Samrat Bora and 
the support of Arya Vidyapeeth College (A), Guwahati.


\end{document}